\documentclass[runningheads]{llncs}
\usepackage[T1]{fontenc}
\usepackage{hyperref}
\usepackage{graphicx}
\usepackage{booktabs}
\usepackage{array}
\usepackage{subfigure}
\usepackage{caption}
\usepackage{float}
\usepackage[misc]{ifsym}
\usepackage{multirow}
\usepackage{amsmath}
\usepackage{mathrsfs}
\usepackage{bbding}
\usepackage{pifont}
\usepackage{threeparttable}
\usepackage{color}

\urldef{\maillcf}\path|chunfeng.lian@mail.xjtu.edu.cn|
\urldef{\mailwf}\path|fan.wang@mail.xjtu.edu.cn|
\urldef{\mailyj}\path|yj1118@mail.xjtu.edu.cn|

\newcommand{\etal}{\emph{et al.}}

\begin{document}
\title{Punctate White Matter Lesion Segmentation in Preterm Infants Powered by Counterfactually Generative Learning}
%
\titlerunning{DeepPWML}
%
 \author{Zehua Ren\inst{1} \and
 Yongheng Sun\inst{2} \and
 Miaomiao Wang\inst{3} \and
 Yuying Feng\inst{3} \and
 Xianjun Li\inst{3} \and
 Chao Jin\inst{3} \and
 Jian Yang\inst{3}\textsuperscript{(\Letter)} \and
 Chunfeng Lian\inst{2}\textsuperscript{(\Letter)} \and
 Fan Wang\inst{1, 3}\textsuperscript{(\Letter)}}
\authorrunning{Z. Ren et al.}
 \institute{ The Key Laboratory of Biomedical Information Engineering of Ministry of Education, School of Life Science and Technology, Xi'an Jiaotong University, China\\
\mailwf\and
  School of Mathematics and Statistics, Xi'an Jiaotong University, China\\
\maillcf\and
 Department of Radiology, The First Affiliated Hospital of Xi'an Jiaotong University, China\\
\mailyj}


%
%
%
\maketitle              
\begin{abstract}
Accurate segmentation of punctate white matter lesions (PWMLs) are fundamental for the timely diagnosis and treatment of related developmental disorders.
Automated PWMLs segmentation from infant brain MR images is challenging, considering that the lesions are typically small and low-contrast, and the number of lesions may dramatically change across subjects.
Existing learning-based methods directly apply general network architectures to this challenging task, which may fail to capture detailed positional information of PWMLs, potentially leading to severe under-segmentations.
In this paper, we propose to leverage the idea of counterfactual reasoning coupled with the auxiliary task of brain tissue segmentation to learn fine-grained positional and morphological representations of PWMLs for accurate localization and segmentation.
A simple and easy-to-implement deep-learning framework (i.e., DeepPWML) is accordingly designed.
It combines the lesion counterfactual map with the tissue probability map to train a lightweight PWML segmentation network, demonstrating state-of-the-art performance on a real-clinical dataset of infant T1w MR images. The code is available at 
\href{https://github.com/ladderlab-xjtu/DeepPWML}{https://github.com/ladderlab-xjtu/DeepPWML}.

\end{abstract}
%
%
%
\section{Introduction}

Punctate white matter lesion (PWML) is a typical type of cerebral white matter injury in preterm infants, potentially leading to psychomotor developmental delay, motor delay, and cerebral palsy without timely treatment~\cite{de2011clinical}.
The early detection and quantitative analysis of PWMLs are critical for diagnosis and treatment, especially considering that some PWML subtypes are only detectable by magnetic resonance imaging (MRI) shortly after birth (e.g., around the third week) and will become invisible thereafter~\cite{debillon2003limitations}.
The PWMLs are small targets that typically locate anterior or adjacent to the ventricles~\cite{kersbergen2014different}.
Manually annotating them in MR images is very time-consuming and relies on expertise.
Thus there is an urgent need, from neuroradiologists, to develop reliable and fully automatic methods for 3D PWML segmentation.

Automated localization and delineation of PWML are practically challenging.
This is mainly because that PWMLs are isolated small objects, with typically only dozens of voxels for a lesion and varying numbers of lesions across different subjects.
Also, due to underlying immature myelination of infant brains~\cite{back2017white}, the tissue-to-tissue and lesion-to-tissue contrasts are both very low, especially in T1w MR images commonly used in clinical practice.
In addition to conventional methods based on thresholding~\cite{cheng2013white} or stochastic likelihood estimation~\cite{cheng2015stochastic}, recent works attempted to apply advanced deep neural networks in the specific task of PWML segmentation~\cite{mukherjee2019fast,liu2019refined,li2021automatic}.
For example, Liu~\etal~\cite{liu2019refined} extended Mask R-CNN~\cite{he2017mask} to detect and segment PWMLs in 2D image slices. Li~\etal~\cite{li2021automatic} implemented a 3D ResU-Net to segment diffuse white matter abnormality from T2w images.
Overall, these existing learning-based methods usually use general network architectures. They may fail to completely capture fine-grained positional information to localize small and low-contrast PWMLs, potentially resulting in high under-segmentations.

Counterfactual reasoning,explained by our task, studies how a real clinical brain image appearance (factual) changes in a hypothetical scenario (whether lesion exist or not). 
This idea has been applied as structural causal models (SCMs) in a deep learning way in recent years. 
At the theoretical level, Monteiro~\etal \cite{monteiro2023measuring} have presented a theoretically grounded framework to evaluate counterfactual inference models.
Due to the advantage of being verifiable, this idea appeared in many medical scenarios.
Pawlowski~\etal \cite{pawlowski2020deep} proposed a general framework for building SCMs and validated on a MNIST-like dataset and a brain MRI dataset. 
Reinhold~\etal \cite{reinhold2021structural} developed a SCM that generates images to show what an MR image would look like if demographic or disease covariates are changed. 
In this paper, we propose a fully automatic deep-learning framework (DeepPWML) that leverages counterfactual reasoning coupled with location information from the brain tissue segmentation to capture fine-grained positional information for PWML localization and segmentation.
Specifically, based on patch-level weak-supervision, we design a counterfactual reasoning strategy to learn voxel-wise residual maps to manipulate the classification labels of input patches (i.e., containing PWMLs or not).
In turn, such fine-grained residual maps could initially capture the spatial locations and morphological patterns of potential lesions. In this article, we define this residual map as counterfactual map which may be different from the meaning of counterfactual map in other articles, hereby declare.
And to refine the information learned by the counterfactual part, we further include brain tissue segmentation as an auxiliary task.
Given the fact that PWMLs have specific spatial correlations with different brain tissues, the segmentation probability maps (and inherent location information) could provide a certain level anatomical contextual information to assist lesion identification. 
Finally, by using the counterfactual maps and segmentation probability maps as the auxiliary input, we learn a lightweight sub-network for PMWL segmentation.

Overall, our DeepPWML is practically easy to implement, as the counterfactual part learns simple but effective linear manipulations, the tissue segmentation part can adopt any off-the-shelf networks, and the PWML segmentation part only needs a lightweight design.
On a real-clinical dataset, our method led to a state-of-the-art performance in the infant PWML segmentation task.

\begin{figure}[t]
\includegraphics[width=\textwidth,height=0.9\textwidth]{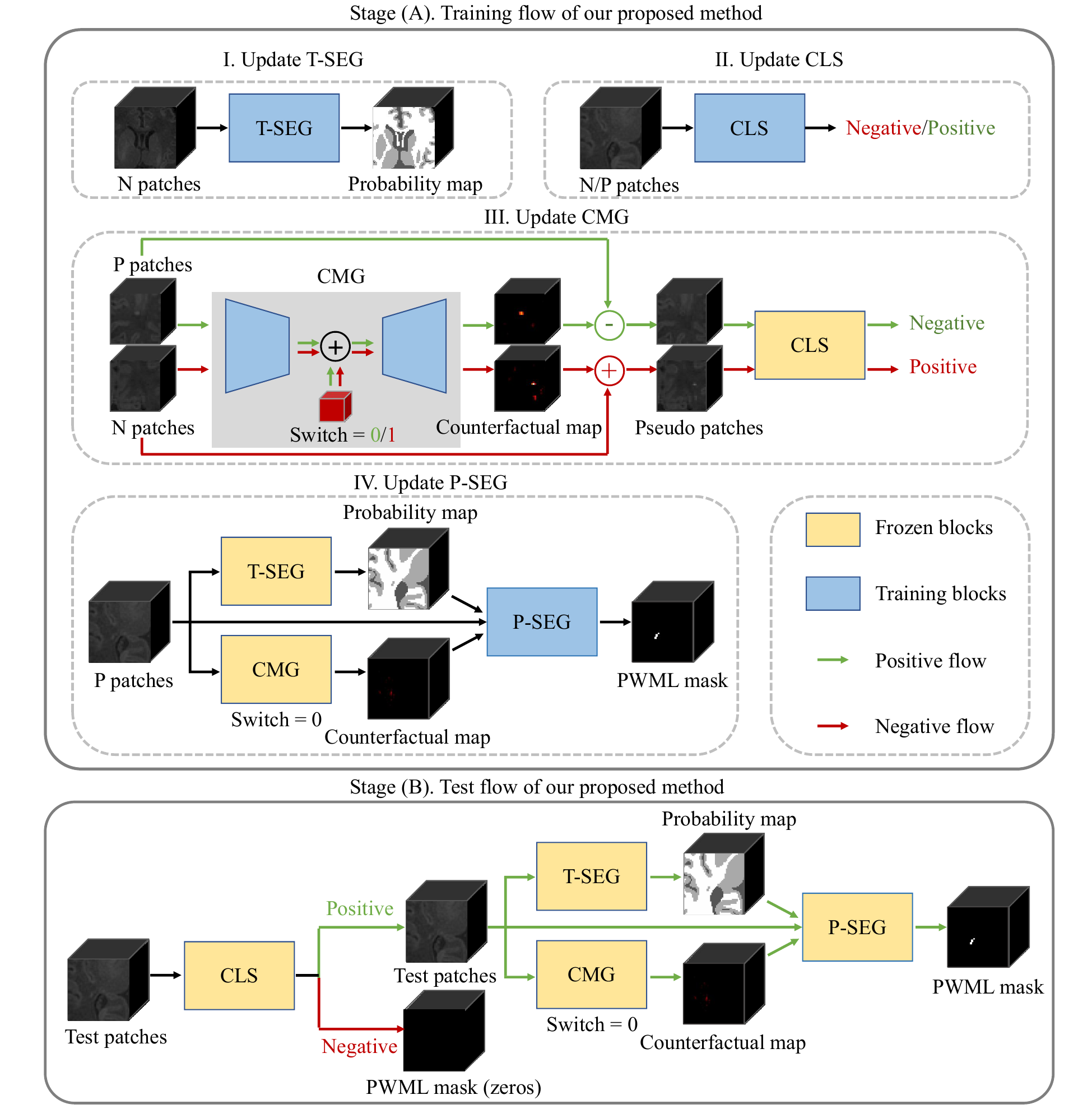}
\caption{Overview of the training and test steps of our DeepPWML framework that consists of four components, i.e., T-SEG, CLS, CMG, and P-SEG modules.}
\label{method}
\end{figure}
\section{Method}
As shown in Fig.~\ref{method}, our DeepPMWL consists of four parts, i.e., the tissue segmentation module (T-SEG), the classification module (CLS), the counterfactual map generator (CMG), and the PWML segmentation module (P-SEG).
Specifically, in the training stage, T-SEG is learned on control data, while other modules are learned on PWML data.
Given an image patch as the input, CLS is trained to distinguish positive (containing PWML) or negative (no PWML) cases, based on which CMG is further trained to produce a counterfactual map to linearly manipulate the input to change the CLS result.

The high-resolution counterfactual maps (CF maps) and segmentation probability maps (SP maps) are further combined with the input patch to train a lightweight P-SEG for PWML segmentation.
In the test stage, an input patch is first determined by the CLS module whether it is positive. Positive inputs will pass through the T-SEG, CMG, and P-SEG modules to get the PWML segmentation results.
It is worth noting that the test patches are generated by sliding windows, and the overlapping results are averaged to get the final segmentation results for the image, which reduces the impact of incorrect classification of the CLS module. 
In our experiments, T-SEG used the voxel-wise Cross-Entropy Loss. CLS used the Categorical Cross-Entropy Loss,  CMG combined the sparsity loss (L1 and L2 norms) with the classification loss. Finally, P-SEG used the Dice Loss.
In the following subsections, we will introduce every module in our design.

\subsection{Tissue Segmentation Module}
 The task is to mark every pixel of the brain as cerebrospinal fluid (CSF), gray matter (GM), or white matter (WM). The choice of this module can be flexible, and there are many off-the-shelf architecture designs available. We adopt a simple Dense-Unet architecture~\cite{zeng20223d} for the T-SEG module. It is trained on control premature infants' images. This module will output the SP map in which segmentation result can be obtained. Therefore, this SP map naturally contains some level anatomy information. Moreover, when an input with PWML goes through a network that has only been trained on control data, the segmentation mistake is partly due to the existence of PWML. Therebefore, this module can output a SP map carrying both potential location and anatomy guidance for PWML localization and segmentation.

\subsection{Classification Module and Counterfactual Map Generator}
The CLS and the CMG are trained sequentially. The CLS is trained to determine whether the current patches have lesions.
The CMG is a counterfactual reasoning step for the CLS. Based on the characteristic of PWML, CMG learns a simple linear sparse transform shown as the CF map. This map aims to offset the bright PWML pixels of the image patches, which are classified as positive, or seed PWML on the patches judged as negative. In other words, CMG is learning a residual activation map for conversion between control and PWML. We adopt the T-SEG module's encoder with two fully connected layers as the CLS module. Furthermore, the architecture of CMG is a simple U-net adding a ``switch'' state in its skip-connection parts according to the method of Oh~\etal~\cite{oh2022learn}. Based on the nature of PWMLs, the last layer of CMG is Relu activation to ensure that the generated CF map is a positive activation. 

The state of the ``switch'' is determined by the classifier's result on the current patch. If the judgement is positive, correspondingly, the ``switch'' status is 0. In this condition, the activated areas in the CF map should be where PWMLs exist. Then the pseudo patches in Fig.~\ref{method}, obtained by subtracting the CF map from the input patches, should be judged as negative by the fixed CLS. Another state of the ``switch'' is used to generate PWMLs. When the CLS judges are negative, the ``switch'' status is 1. in this situation, the input patches combining the CF map should be classified as positive. This training strategy is to make CMG learn PWML features better. When it comes to the test phase, the switch status will be fixed to 0. Because in the test phase, the CF map only needs to capture PWML.


The CMG module is summarised as follows:
Firstly, PWML patches $C_P$ and control patches $C_N$ are fed to the encoder to obtain encoded representations $F_P$ and $F_N$:
\begin{align}
        F_P &= {\rm Encoder}(C_P), \\
        F_N &= {\rm Encoder}(C_N).
\end{align}
Secondly, ``switch'' filled with zeros/ones with the same size as PWML/normal representations $F_P$/$F_N$ are added to these representations and then pass through the decoder to obtain the CF maps $M_P$/$M_N$:
\begin{align}
        M_P &= {\rm Decoder} (F_P + Zeros), \\
        M_N &= {\rm Decoder} (F_N + Ones).
\end{align}
Finally, the original patches $C_P$/$C_N$ are added/subtracted to the CF maps $M_P$/$M_N$ to yield the transformed patches $\widetilde C_P$/$\widetilde C_P$, which are classified by the CLS module as the opposite classes:
\begin{align}
        \widetilde C_P &= C_P + M_P, \\
        \widetilde C_N &= C_N - M_N.
\end{align}

\subsection{PWML Segmentation Module}
The SP map includes the potential PWML existence, but also a lot of tissue segmentation uncertainty. The CF map directly shows the PWML location, but due to the accuracy of the CLS module, the CF map itself will also carry some false positives fault. If we synthesize the CF map, the SP map and the original input patches for appearance information, the best segmentation result can be achieved by allowing the network to verify and filter out each information in a learnable way.

The P-SEG module is implemented as a lightweight variant of the Dense-Unet. Different simplified versions have been tested, with the results summarized in Section~\ref{subsec:results}.
After getting the PWML segmentation result, we use the tissue segmentation result to filter out PWMLs mis-segmented at the background and CSF.

\section{Experiments and Results}
\subsection{Dataset and Experimental Setting}
\noindent\textbf{Dataset:}
Experiments were performed on a dataset with two groups (control and PWML), where control included 52 subjects without PWML observed, and PWML included 47 subjects with PWMLs. All infants in this study were born with gestational age (GA) between 28 to 40 weeks and scanned at post-menstrual age (PMA) between 37 to 42 weeks. Two neuroscientists manually labeled PWML areas and corrected tissue labels generated by iBeat~\cite{dai2013ibeat}. Written Consent was obtained from all parents under the institutional review board, and T1-weighted MR images were collected using a 3T MRI scanner, resampling the resolution of all images into 0.9375 $\times$ 0.9375 $\times$ 1 mm$^3$. All images are cropped to 130 $\times$ 130 $\times$ 170.

\noindent\textbf{Experimental Setting:}
 Our method was implemented using Tensorflow. All modules were trained and tested on an NVIDIA GeForce RTX 3060 GPU. We adopted Adam as the optimizer, with the learning rate varying from 0.001 to 0.00001 according to modules. The inputs were fixed-size patches ($32\times32\times32$) cut from the T1w images. The train/validation/test ratio was 0.7/0.15/0.15 and divided on subject-level. We didn’t use any data augmentation during training. We used Dice, True Positive Rate (TPR), and Positive Predictive Value (PPV) to quantitatively evaluate the segmentation performance.

\begin{figure}[t]
\includegraphics[width=\textwidth]{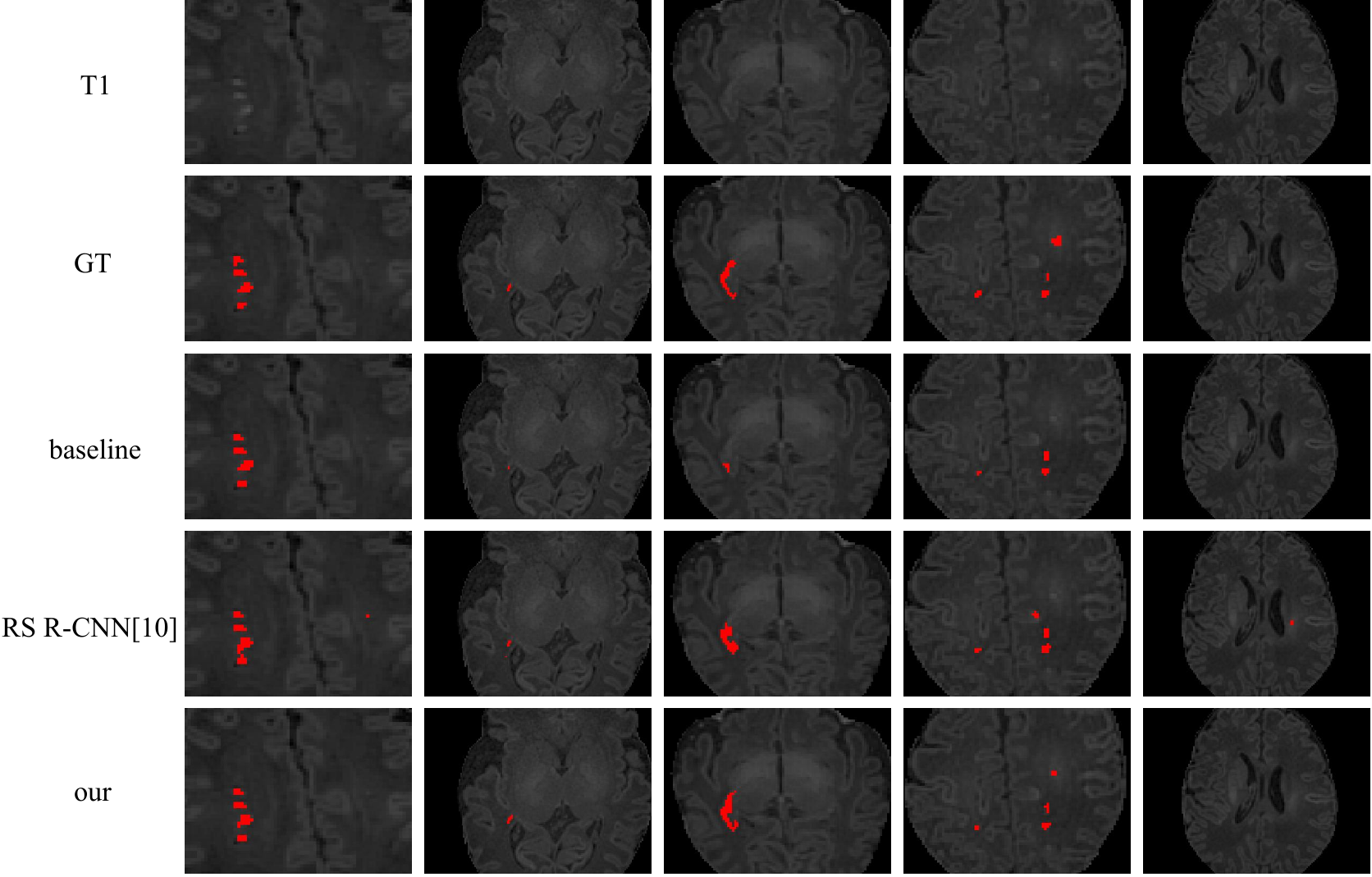}
\caption{Visual comparisons of the representative PWML segmentation results.}
\label{comp}
\end{figure}

\subsection{Results} \label{subsec:results}
First, the T-SEG module is trained using a fully supervised way. Its tissue segmentation accuracy on the test set is about 93\% in terms of Dice. Second, the CLS and other modules are trained with PWML group data. We defined the input training patches' class labels by whether they contain PWMLs or not. In other words, if any patch has at least one lesion voxel, it is positive. The accuracy of the test set can reach around 90\%. Third, we train the CMG module based on the well-trained and fixed CLS module. Finally, based on T-SEG and CMG, we train P-SEG. We combine the SP map, CF map, and T1w image in a channel-wise way as the input of the module without any additional processing of these features.

\noindent\textbf{Comparison Results:}
We compared our method with the state-of-the-art method~\cite{liu2019refined}. As is shown in Table~\ref{tab:comp2}, our method outperforms the state-of-the-art method and the baseline model in all three indexes. The visualization results are shown in Fig.~\ref{comp}, from which it can be seen that our method can segment small-size PWMLs more accurately and segment PWMLs with different severities more completely. 

\begin{table}[t]
\setlength{\tabcolsep}{6pt}
\centering
\caption{Comparison of our method (and its variants) with the state-of-the-art method and the baseline model. All metrics are presented as “mean (std)”.}
\label{tab:comp2}
\begin{tabular}{cc|ccc}
\hline
\multicolumn{2}{c|}{Methods}                                       & Dice            & TPR             & PPV             \\ \hline
\multicolumn{2}{c|}{Baseline~\cite{zeng20223d}}                                      & 0.649(0.239)          & 0.655(0.244)          & 0.704(0.281)          \\
\multicolumn{2}{c|}{RS R-CNN~\cite{liu2019refined}}            & 0.667(0.172)          & 0.754(0.250)          & 0.704(0.187)               \\ \hline
\multicolumn{1}{c|}{\multirow{6}{*}{Ours}} & SP map           & 0.649(0.142)          & 0.726(0.210)          & 0.677(0.213)          \\
\multicolumn{1}{c|}{}                      & CF map                 & 0.507(0.169)          & 0.543(0.286)          & 0.647(0.180)         \\
\multicolumn{1}{c|}{}                      & SP map + T1        & 0.680(0.178)          & 0.794(0.211)          & 0.699(0.237)          \\
\multicolumn{1}{c|}{}                      & CF map + T1              & 0.672(0.198)          & 0.741(0.249)          & 0.719(0.178)          \\
\multicolumn{1}{c|}{}                      & SP map + CF map    & 0.670(0.184)          & 0.781(0.181)          & 0.684(0.251)          \\
\multicolumn{1}{c|}{}                      & SP map + CF map + T1 & \textbf{0.721(0.177)} & \textbf{0.797(0.185)} & \textbf{0.734(0.211)} \\ \hline
\end{tabular}
\end{table}

\begin{figure}[t]
  \centering
\includegraphics[width=\textwidth]{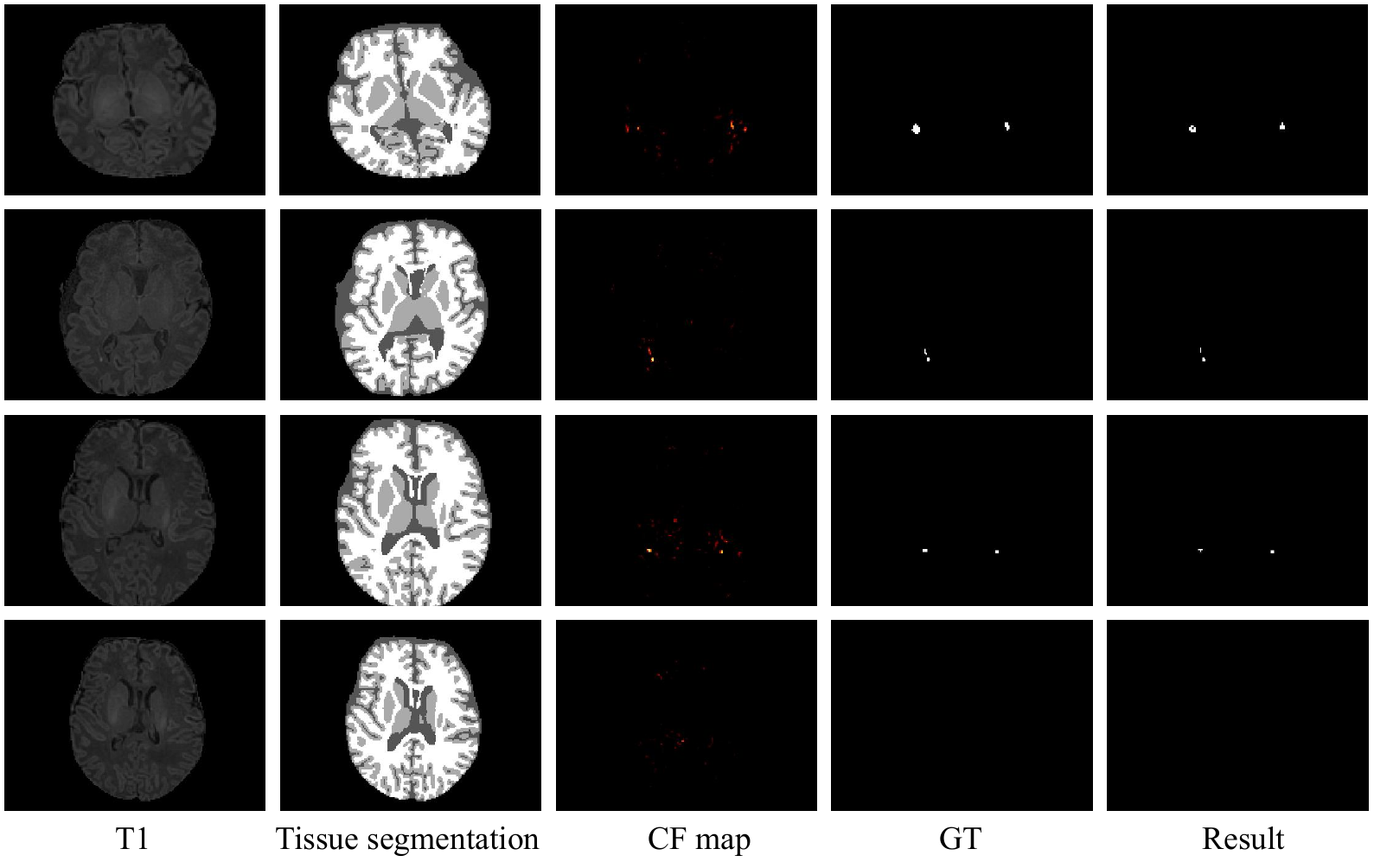}
\caption{Example visualization of T1w images, tissue segmentation maps, CF maps, labels, and segmentation results. Tissue segmentation maps are the final segmentation output of the T-SEG module. CF maps are the output of the CMG module.}
\label{maps}
\end{figure}

\noindent\textbf{Ablation Studies:}
We further evaluated the effectiveness of our design by comparing the results of the pipelines with and without SP maps and CF maps. The ablation results are shown in the last six rows of Table~\ref{tab:comp2}. The baseline model using the same dense-Unet is trained to segment the PWML from T1w images. Other settings are consistent with our final module. Then we will add our designed modules step by step to verify the effectiveness of two kinds of auxiliary information.

 By comparing ``baseline'', ``SP map'', and ``CF map'', we can find that the two kinds of information individually are not good for segmenting PWMLs. The reason is as follows. The SP map mainly focuses on tissue segmentation task. The CF map has some false activation due to the offset of the highlighted areas for PWML. Fusing these two kinds of information has reduced their respective defects (``SP map + CF map'').  The icing on the cake is that when the appearance features of T1w are used again, the accuracy will be significantly improved (``SP map + T1'' and ``CF map + T1''). This means ``SP map'' and ``CF map'' each can be an auxiliary information but not sufficient resource for this task. Finally, after combining the three together, all indicators have been significantly improved (``SP map + CF map + T1'').

\noindent\textbf{Visual Analysis:}
Fig.~\ref{maps} shows the T1w images, tissue segmentation maps, CF maps, labels, and segmentation results. By selecting the most likely category from the SP map as the label, the tissue segmentation map can be obtained. As shown in the tissue segmentation maps, PWML voxels tend to be classified as gray matter surrounded by white matter which obviously does not conform to the general anatomy knowledge. The reason of this phenomenon may be that the intensity of gray matter and PWML are higher than white matter in T1w image at this age. It also can be seen from the CF maps that these maps have a preliminary localization of PWML. The last row shows the situation without PWML. It can be seen that the tissue segmentation is reasonable. The CF map has a small amount of activation and the intensity is significantly lower than the first three rows. 
In conclusion, these two maps complementarily provide the anatomical and morphological information for the segmentation of the PWML.

\noindent\textbf{Comparison of different backbones of the P-SEG Module:}
We test from simple several layers to the whole dense-Unet to determine the required complexity in Table~\ref{tab:comp}. We compared six designs with different network sizes in Table~\ref{tab:comp}. The first three methods are several convolution layers with the same resolution. The latter three reduce the number of down-samplings in the original dense-Unet. By comparing the Dice index, it is obvious that the simple convolution operation cannot integrate the three kinds of input information well. The results show that the encoder-decoder can better fuse information. Perhaps because of the small size of PWML, it does not require too much down-sampling to get a similar result as the optimal result. The result also indicates that a certain degree of network size is needed to learn the PWMl characteristics.

\begin{table}[t]
\setlength{\tabcolsep}{6pt}
\centering
\caption{Comparison of different backbones of the P-SEG module.}
\label{tab:comp}
\begin{tabular}{@{}lcc@{}}
\toprule
Methods                            & Dice   & Network size (KB) \\ \midrule
Four convolutional layers            & 0.5843 & 107               \\
One Dense-Block                    & 0.6220 & 420               \\
Two Dense-Blocks                   & 0.6644 & 836               \\
Dense-Unet with one down-sampling  & 0.7099 & 2127              \\
Dense-Unet with two down-sampling & 0.7180 & 5345              \\
Dense-Unet                         & \textbf{0.7214} & 9736              \\ \bottomrule
\end{tabular}
\end{table}

\section{Conclusion}
In this study, we designed a simple and easy-to-implement deep learning framework (i.e. DeepPWML) to segment PWMLs. Leveraging the idea of generative counterfactual inference combined with an auxiliary task of brain tissue segmentation, we learn fine-grained positional and morphological representations of PWMLs to achieve accurate localization and segmentation. Our lightweight PWML segmentation network combines lesion counterfactual maps with tissue segmentation probability maps, achieving state-of-the-art performance on a real clinical dataset of infant T1w MR images. Moreover, our method provides a new perspective for the small-size segmentation task.
%
%
%
\bibliographystyle{splncs04}
\bibliography{paper935}
%




\end{document}